# Capacitor-Type Thin-Film Heat Flow Switching Device

Keisuke Hirata (1, 2), Takuya Matsunaga (1, 2), Saurabh Singh (1), Masaharu Matsunami (1), and Tsunehiro Takeuchi (1)

(1) Toyota Technological Institute, 2-12-1 Hisakata, Tempaku-Ku, Nagoya 468-8511, Japan

(2) Corresponding author

E-mail: sd20502@toyota-ti.ac.jp

- The authors declare that they have no conflict of interest.
- The data that support the findings of this study are available from the corresponding author upon reasonable request.

**Abstract**

We developed a capacitor-type heat flow switching device, in which electron thermal conductivity of the electrodes is actively controlled through the carrier concentration varied by an applied bias voltage. The devices consist of an amorphous p-type Si-Ge-Au alloy layer, an amorphous $SiO_2$ as the dielectric layer, and a n-type Si substrate. Both amorphous materials are characterized by very low lattice thermal conductivity, $\leq 1$ $Wm^{-1}K^{-1}$. The Si-Ge-Au amorphous layer with 40 nm in thickness was deposited by means of molecular beam deposition technique on the 100 nm thick $SiO_2$ layer formed at the top surface of Si substrate. Bias voltage-dependent thermal conductivity and heat flow density of the fabricated device were evaluated by a time-domain thermoreflectance method at room temperature. Consequently, we observed a 55% increase in thermal conductivity.

―――――――――――――――――

**Introduction**

The concept of "*heat management*" describes the effective ways of using the vast amount of waste heat generally abandoned in an environment. It contains technologies and techniques such as thermal storage [1, 2], thermal insulator [3], thermal diode [4–8], thermal switch [9–17], etc [18]. Once we could develop a heat flow switching device with which one can actively tune the magnitude and direction of heat flow in the device, it would attract considerable interest as one of the key technologies in heat management. In this study, we tried to develop such a new heat flow switching device using thin films.

Fourier's law, $\boldsymbol{J_Q} = -\kappa \nabla T$, tells us that heat flow density $\boldsymbol{J}_\mathrm{Q}$ in isotropic solids is determined simply by its thermal conductivity $\kappa$ and a temperature gradient $\nabla T$. Under a given temperature gradient $\nabla T$, therefore, a variation in thermal conductivity should be required to change the magnitude of the heat flow density. In solid materials, conduction electrons together with collective excitations, such as phonon (lattice vibration), magnon, and spinon, contribute to thermal conductivity. The contribution of these factors to thermal conductivity is expressed as $\kappa_\mathrm{el}$, $\kappa_\mathrm{lat}$, $\kappa_\mathrm{mag}$, and $\kappa_\mathrm{sp}$, respectively, and the sum of these partial thermal conductivity determines the total thermal conductivity $\kappa$. Therefore, at least one of them should be dynamically controlled to develop heat flow switching devices.

In the past decade, various mechanisms were proposed to realize a finite variation in heat flow density through a variation in thermal conductivity; (a) structure modulation or phase transition derived change in $\kappa_{el}$ and $\kappa_{lat}$ [9–13], (b) domain structure control of ferroelectric materials leading to a variation in $\kappa_{lat}$ [14, 15], (c) spin chain ladder cuprates in which magnon transport varies with $\kappa_{mag}$ [16], (d) carrier concentration tuning with bias voltage for varying $\kappa_{el}$ [17]. However, none of them has reached any practical applications, mainly because their performance was not good enough. We also consider that, in addition to the large variation in magnitude of heat flow density, an ideal practical device should have some more features such as quick response, controllability, simple design, and ubiquitous nontoxic constituent elements.

In order to develop such a device, in our recent study [17], we fabricated a capacitor-type heat flow switching device working with an applied bias voltage. Figure 1 (a) schematically illustrates the basic structure of our proposing device consisting of p- and n-type semiconductor electrodes separated by a dielectric layer. When a forward voltage is applied to the device, as shown in Fig. 1 (b), the carriers (electrons/holes) are introduced into the electrodes to increase their carrier density near the layer-boundaries. The increased number of carriers naturally leads to an increase in electrical conductivity and hence electron thermal conductivity. The time required for changing carrier concentration is expected to be the same as that of a general electric capacitor [19].

We should note here that very small lattice thermal conductivity is required for both the insulating dielectric material and electrode material in the device because the constant magnitude of lattice thermal conductivity naturally reduces the variation of total thermal conductivity $\kappa_{tot}$ w/ or w/o a bias voltage $V_B$ as described in the following equation.

$$\frac{\kappa_{tot}(V_B)}{\kappa_{tot}(0)} = \frac{\kappa_{el}(V_B) + \kappa_{lat}}{\kappa_{el}(0) + \kappa_{lat}} \tag{1}$$

Equation (1) suggests that we cannot observe any effective change in $\kappa_{tot}$ when $\kappa_{lat}$ is much larger than the difference between $\kappa_{el}(V_B)$ and $\kappa_{el}(0)$. For obtaining an effective change in thermal conductivity, therefore, our strategy was to employ degenerate semiconductors possessing an extremely low $\kappa_{lat}$ as the electrode materials. Besides, for the significant changes in $\kappa_{el}$ by tuning carrier concentration, both the electronic structure of degenerate semiconductors and moderate electrical conductivity for easy and quick carrier injection are required.

In our first development of device [17], n-type semiconductor $Ag_2S_{0.6}Se_{0.4}$ was selected as one of the electrodes because of its very low lattice thermal conductivity (0.3 – 0.4 Wm$^{-1}$K$^{-1}$) and combined with non-doped amorphous Si as the dielectric layer. This device successfully led to the linearly increasing heat flow density with bias voltage and achieved a 10% increase with a small voltage of $V_B$ = +3 V. The total thickness of the device was about 10 $\mu$m, while the estimated thickness of the layer contributing to the variation in thermal conductivity was supposed to be limited to a few tens of nanometer, which corresponds to only about 0.1% of the total device-thickness. This experimental fact strongly suggested that an enhanced performance would be obtainable by thinning electrodes and a dielectric layer.

Therefore, in this study, we developed a new capacitor-type heat flow switching device composed of thin layers for improving their switching performance and reported the successfully obtained enhanced variation in thermal conductivity and heat flow density through the device.

───────────────────────

**Experimental Procedure**

A capacitor-type heat flow switching device was constructed on the n-type Si substrate that works as one of the electrodes. The electrically and thermally insulating amorphous SiO$_2$ dielectric layer (9 eV bandgap [20] and 1 Wm$^{-1}$K$^{-1}$ in thermal conductivity) of 100 nm in thickness was formed at the top surface of the substrate by thermal oxidation at 1173 K under H$_2$ and O$_2$ mixed gas-flow. The other electrode of p-type amorphous Si-Ge-Au alloys was selected due to its low thermal conductivity [21] (0.7 – 2 Wm$^{-1}$K$^{-1}$) and deposited with 40 nm in thickness on the SiO$_2$ layer. We employed molecular beam deposition (MBD) apparatus developed by EIKO Engineering for making Si-Ge-Au layers using three crucibles, each of which contains one of Si (5N), Ge (4N), and Au (3N). These constituent elements were co-evaporated towards the substrate, which was not heated up but slightly warmed by the radiation from the element source. The pressure of the MBD chamber was 5×10$^{-8}$ Pa in the base condition but slightly increased to 10$^{-5}$ – 10$^{-7}$ Pa during the depositions.

At the top surface of each device, we further deposited the Mo layer by means of radiofrequency magnetron sputtering using ULVAC, VTR-150M/SRF. The Mo layer works as the other electrode of the capacitor-type device for applying electrical field homogeneously to the Si-Ge-Au layer. It also works as the reflector of 755 nm (1.6 eV) laser and the absorber of 1550 nm (0.8 eV) laser in the time-domain thermoreflectance (TDTR) measurements.

Previously reported data for the thermal conductivity of amorphous Si-Ge thin film [21] suggested that the lowest lattice thermal conductivity ~ 0.7 Wm$^{-1}$K$^{-1}$ is obtainable at the vicinity of Si$_{40}$Ge$_{60}$ composition. Therefore, we selected this ratio for the base composition. In order to make electrical conductivity moderately large at around 1 – 100 Scm$^{-1}$, we used Au as a p-type dopant. Thus, Si$_{40-0.5x}$Ge$_{60-0.5x}$Au$_x$ ($x$ = 3, 5, 15, and 20 at.%) were prepared, and their electrical conductivity was measured to determine the best composition for the electrode of capacitor-type heat flow switching device.

The crystallographic analyses for thin films were performed using the X-ray diffraction (XRD) with Cu-$K\alpha$ ($\lambda$ = 1.5418 Å) line in Bruker, D8 ADVANCE. The surface structure, homogeneity, and composition of the thin films were confirmed by scanning electron microscope (SEM) and energy-dispersive X-ray spectroscopy (EDX) with 20 keV accelerating-voltage using Hitachi, SU6600.

The room temperature electrical conductivity of Si-Ge-Au thin films was measured by the conventional four-probe method. The absolute value of electrical conductivity was determined from the slope of current-voltage relation with a small electrical current from 1 to 10 $\mu$A to avoid degradation of the thin-film samples. The thickness of each layer was determined by a step profiler (KLA, ALPHA STEP IQ), and the width and inter-probe distance for voltage measurements were measured by a micrometer with the maximum error of 0.1%.

The picosecond-range TDTR apparatus [22] (PicoTherm, Pico-TR) was used to investigate the thermal conductivity of the $Si_{40-0.5x}Ge_{60-0.5x}Au_x$ ($x$ = 3, 5, 15, and 20 at.%) and the "*thermal conductivity of the heat flow switching device*" at room temperature. Here, we evaluated the "*thermal conductivity of the device*" with hypothetically assuming that the device is made up of "*single, homogeneous material*". The schematic illustration of the device is shown in Fig. 2 together with the experimental setup of the TDTR method. The bias voltage-dependent "*thermal conductivity of the device*" in the stacking direction was measured by the front-heating front-detection method with monitoring a leak current through the device. The specific heat and density of Mo and Si-Ge-Au thin film were considered to be the same as those of bulk samples [21].

---

**Results**

Figure 3 shows XRD patterns of $Si_{40-0.5x}Ge_{60-0.5x}Au_x$ ($x$ = 3, 5, 15, and 20 at.%) thin films (100 nm in thickness) deposited on a glass substrate. For the samples prepared at $x$ = 3 and 5 at.%, we observed a halo-pattern indicating the formation of an amorphous phase without having any recognizable peaks from precipitated crystals. In the samples prepared at $x$ = 15 and 20 at.%, three XRD peaks appeared at $2\theta$ = 28, 38, and 46° together with the halo-background of amorphous phase. These peaks were attributed to 111 of diamond structure Si-Ge, 111 of fcc Au, and 200 of Si-Ge, respectively. The mean diameter of crystalline particles was estimated from the full width at half maximum of each XRD peaks using Scherrer's equation [23], and the results were 15 nm for Au and 20 – 30 nm for Si-Ge, respectively. The precipitation of Au and Si-Ge nano-crystals at higher Au concentrations is consistent with previous studies [21, 24].

The formation of a homogeneous amorphous phase for $Si_{37.5}Ge_{57.5}Au_5$ was confirmed by the secondary electron SEM imaging [Fig. 4(a)] and elemental distribution EDX mapping [Fig. 4(b-d)]. There were no noticeable voids nor cracks on the surface at least in the measured area. With multiple scans, we safely confirmed that the compositions determined by EDX were almost the same as the nominal ones within an error of a two at.%.

Figure 5(a) shows the Au concentration dependence of electrical conductivity observed for $Si_{40-0.5x}Ge_{60-0.5x}Au_x$ ($x$ = 3, 5, 15, and 20 at.%). The electrical conductivity obviously increased with increasing $x$. The increasing ratio of electrical conductivity to Au concentration was clearly different between the samples at $x \leq 5$ at.% and $x > 5$ at.%. Since nano-crystals were found to precipitate at $x \geq 15$ at.%, this difference should be attributed to the presence of nano-crystals at $x \geq 15$ at.% and its absence at $x \leq 5$ at.%

The composition-dependent thermal conductivity of $Si_{40-0.5x}Ge_{60-0.5x}Au_x$ ($x$ = 3, 5, 15, and 20 at.%) are shown in Fig. 5(b). The thermal conductivity stayed at a very low value below 1 Wm$^{-1}$K$^{-1}$ at $x \leq 15$ at.% and showed an increasing tendency with increasing $x$. We roughly estimated the contribution of the electron thermal conductivity ($\kappa_{el}$) using the Wiedemann–Franz law: $\kappa_{el} = L_0 \sigma T$, where $L_0 = \frac{\pi^2 k_B^2}{3e^2} = 2.44 \times 10^{-8}$ W$\Omega$K$^{-2}$ represents the constant known as Lorenz number. Here, $k_B$ and $e$ are Boltzmann constant ($1.38 \times 10^{-23}$ J K$^{-1}$) and elementary charge of the electron ($-1.6 \times 10^{-19}$ C), respectively. The lattice thermal conductivity ($\kappa_{lat}$) was also roughly estimated by subtracting $\kappa_{el}$ from the measured thermal conductivity ($\kappa$) as $\kappa_{lat} = \kappa - \kappa_{el}$. Obviously, the lattice thermal conductivity was dominant at $x \leq 15$ at.%. The moderately increasing tendency of $\kappa_{lat}$ with increasing $x$ would be related to the local phonon modes of Au that should exist in low energies due to the heavy mass and weaker chemical bonds to be easily excited even at room temperature. The thermal conductivity of $x$ = 20 at.% definitely possessed a much larger magnitude than that in $x \leq 15$ at.% partly because the $\kappa_{el}$ contribution of the amorphous phase became large to be comparable with $\kappa_{lat}$, and partly because the precipitated crystals possess much higher thermal conductivity than that of the amorphous phase.

We considered that the single-phase amorphous $Si_{37.5}Ge_{57.5}Au_5$ possessing the low lattice thermal conductivity (~ 0.7 $Wm^{-1}K^{-1}$) and moderate electrical conductivity (8.2 $Scm^{-1}$) is suitable for the p-type electrode in a capacitor-type heat flow switching device. Figure 2 shows a schematic illustration of the fabricated heat flow switching device consisting of p-type amorphous $Si_{37.5}Ge_{57.5}Au_5$ / amorphous $SiO_2$ / n-type Si substrate. The TDTR signals with a bias voltage of +50 and –17 V after a pulse heating are shown in Fig. 6(a). The data represents the transient surface temperature. The pulse laser was shed on the surface of the device at $t = 0$ ns to instantly increase the temperature, and the surface temperature gradually decreased because of the heat conduction from the surface to the bottom of the device. The cooling rate of the surface temperature clearly varied depending on the applied bias voltage. The fast cooling-rate implies that a larger heat flow density is generated with the forward voltage, while the slow rate at the negative bias indicates a rather resistive nature for the heat flow. These results enabled us to confirm that the variation in carrier concentration leads to an effective change in the total heat flow density through the device.

The relaxation process ($t \leq 10$ ns) was used to estimate the "*thermal conductivity of the device*" that was mainly determined by the Si-Ge-Au layer adjacent to the Mo layer. The "*thermal conductivity of device*" obtained by the fitting of transient surface temperature was plotted as a function of bias voltage and shown in Fig. 6(b) together with the bias voltage dependence of leak current. At low bias voltage, the estimated "*thermal conductivity of device*" increased linearly with increasing voltage. The maximum increase in "*thermal conductivity of device*" was reached to 55% from $V_B = -17$ to $+50$ V.

From the zero-bias voltage, the thermal conductivity increased with forwarding bias and decreased with the reverse case. The former phenomenon should be caused by the principle shown in Fig. 1(a), while the latter phenomenon was most likely due to the mechanism that reverse bias extracted carriers and formed a depletion layer near the boundary between Mo and Si-Ge-Au.

The estimated "*thermal conductivity of the device*", however, turns out to become comparable with the zero-bias value at a higher voltage exceeding +50 V, where a leak current was clearly observed. The leak current naturally reduces the accumulated carriers at the boundaries, and it should be the reason for the disappearance of variation in thermal conductivity. Notably, this tendency was also observed in the reverse bias voltage. It is stressed, from this fact, that the insulating properties of the dielectric layer are another important factor for improving heat flow switching performance.

We considered that the information about variation in heat flow density could be directly obtained from the raw data even without using the complicated function fitting on the transient curve of surface temperature. For example, in Fig. 6(a), we focused on the time required for the temperature change of 20 % from the radiation of the heating laser. It took 26 ns when $V_B = +50$ V, while 47 ns when $V_B = -17$ V. From this difference of time, we roughly estimated (47 – 26)×100 / 47 ~ 45% change also in the heat flow density throughout the device. This number shows good consistency with the variation of "*thermal conductivity of the device*" obtained by the function fitting. We stress here that the performance of our newly developed heat flow switching device is not good enough yet but strongly indicates the potential of our proposing mechanism to realize practical heat flow switching devices.

## Discussion

To establish a quantitative criterion for the thinning of electrodes, from the experiment shown in Fig. 2 and Fig. 6, we roughly calculated the thickness of the layer contributing to the variation in thermal conductivity (in carrier concentration) by assuming the variation of "*thermal conductivity of the device*" occurred in only Si-Ge-Au layer.

The variation in carrier concentration due to the bias voltage $\Delta n_B$ is expressed as $\Delta n_B = \frac{Q}{St|e|}$. Here, $t$ and $e$ are the thickness of the carrier-accumulated layer near the interface and the elementary charge of the electron ($-1.6 \times 10^{-19}$ C), respectively. Since the electron thermal conductivity variation with $V_B$ = 50 V was 30% from zero-bias, we could get the next relationship from Wiedemann–Franz law and Drude model $(\sigma = \frac{ne^2\tau}{m})$.

$$\frac{\Delta\kappa_{el}(50V)}{\kappa_{el}(0V)} = \frac{L_0 \Delta\sigma_B T}{L_0 \sigma T} = \frac{\Delta n_B e^2 \tau}{m} \bigg/ \frac{ne^2\tau}{m} = \frac{Q}{St|e|n} = 0.3 \quad (2)$$

Here, $\sigma$, $n$, $\tau$, and $m$ represent electrical conductivity, the carrier concentration of Si-Ge-Au electrode without bias voltage, relaxation time, and electron mass, respectively. By applying $n = 5 \times 10^{19}$ cm$^{-3}$ deduced by Hall effect measurement to Eq.(2), we roughly estimated the thickness of a carrier-accumulated layer at $t = 8$ nm. This thickness is only 20% of the Si-Ge-Au electrode. Our estimation was certainly rough; however, the experimentally confirmed ~ 55% variation in total electron thermal conductivity implies that a 275% switching ratio at the maximum would be obtainable by reducing an electrode-thickness to 8 nm.

Before finishing discussions, we propose here three additional strategies for realizing a larger switching ratio in heat flow density.

The first one is employing materials possessing smaller lattice thermal conductivity. The electrode should be replaced with the material whose lattice thermal conductivity is smaller than that of amorphous Si-Ge alloys. We have already found and proposed $Ag_2S_{0.6}Se_{0.4}$ as a possible candidate with an extremely low lattice thermal conductivity of $0.3 - 0.4$ $Wm^{-1}K^{-1}$ [5, 17], which is almost half the value of amorphous Si-Ge alloys.

The second strategy is related to the leak current through the dielectric layer. We found that the leak current extinguished the proportional variation in thermal conductivity (heat flow density) to a bias voltage. Therefore, we should make the threshold voltage of current leakage as higher as possible. The preparation of a high-quality insulating layer is definitely one of the solutions. The use of insulating materials with a high relative dielectric constant, such as $HfO_2$ possessing $\varepsilon_r = 25$ [20] is another solution.

The preparation of multilayer is the third strategy. Since the carrier accumulation occurs only in the vicinity of the layer-boundary, an increased number of interfaces must contribute to a larger variation in heat flow density through the device.

---

**Conclusion**

In this study, we fabricated a thin-film capacitor-type heat flow switching device. To select the suitable materials for the electrodes of the device, we investigated the heat-transport properties of p-type amorphous $Si_{40-x}Ge_{60-x}Au_x$ ($x$ = 3, 5, 25, and 20 at.%) at room temperature. As a result, amorphous $Si_{37.5}Ge_{57.5}Au_5$ possessing a very low lattice thermal conductivity (~ 0.7 Wm$^{-1}$K$^{-1}$) and a moderate electrical conductivity (~ 8 Scm$^{-1}$) was selected as one of the suitable materials. A new device was made with a p-type $Si_{37.5}Ge_{57.5}Au_5$ electrode layer (40 nm in thickness), $SiO_2$ glass dielectric layer (100 nm), and n-type Si substrate as the other electrode. The performance of the device was evaluated by the TDTR method w/ or w/o the bias voltage. Consequently, we confirmed that the maximum thermal conductivity variation of 55% with ~ 70 V bias voltage. From this result, we succeeded in demonstrating the validity of our new strategy for developing an enhanced heat flow switching device.

**Figures and Captions**

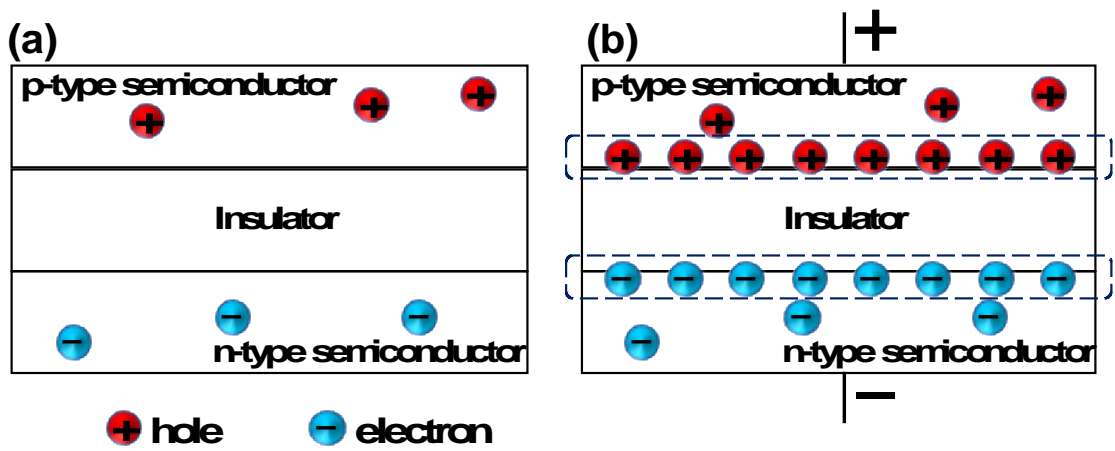

Fig. 1    Schematic illustrations of the principle of a capacitor-type heat flow switching device. (a) shows a typical unit structure of the device consisting of p-type, n-type semiconductors, and an insulator. When a bias voltage was applied, carriers (electrons and holes) are introduced in the electrodes near the layer-boundary, as shown in (b), leading to an increased electron thermal conductivity and heat flow density.

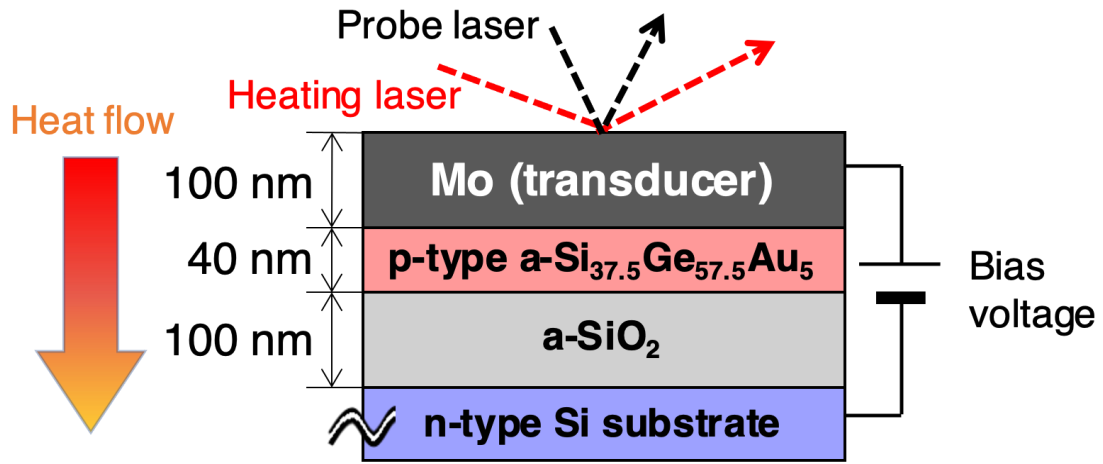

Fig. 2  The schematic layout of a capacitor-type heat flow switching device consisting of amorphous p-type $Si_{37.5}Ge_{57.5}Au_5$, amorphous $SiO_2$, and n-type Si substrate and an arrangement for time-domain thermoreflectance measurement with a bias voltage.

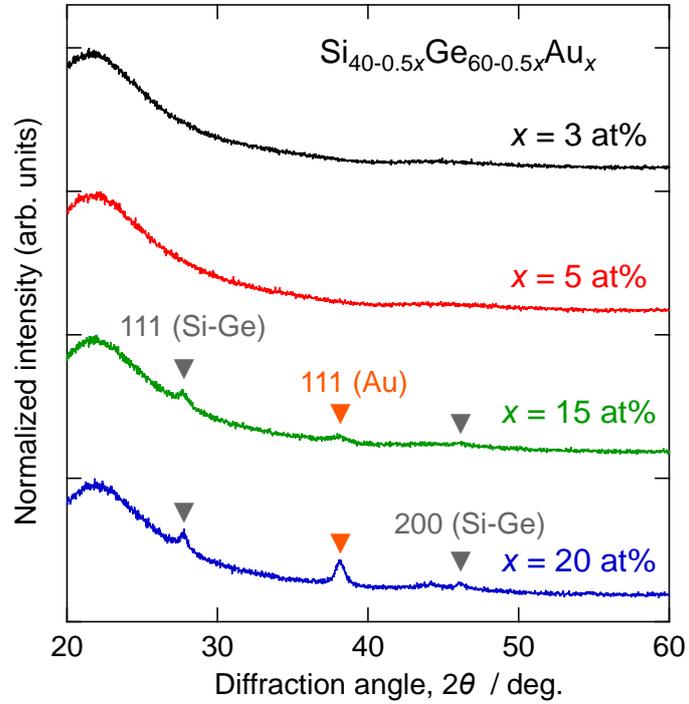

Fig. 3  XRD patterns for $Si_{40-x}Ge_{60-x}Au_x$ ($x$ = 3, 5, 25, and 20 at%) thin-film measured with $\theta - 2\theta$ method at room temperature. The halo patterns imply the formation of the amorphous Si-Ge-Au alloy, whereas several peaks were indexed to the Si-Ge or Au crystals.

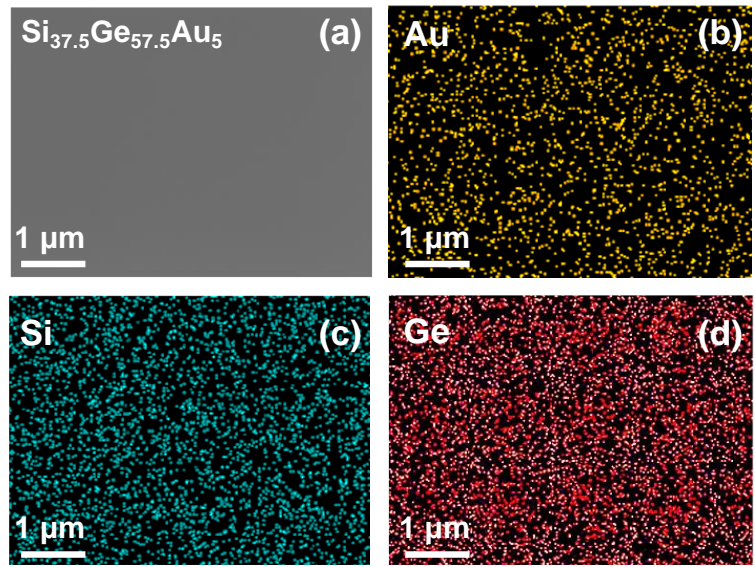

Fig. 4　SEM-EDX observation for amorphous $Si_{37.5}Ge_{57.5}Au_5$ thin film. (a) is a secondary electron imaging. (b), (c), and (d) represent elemental distribution mapping for Au, Si, and Ge, respectively.

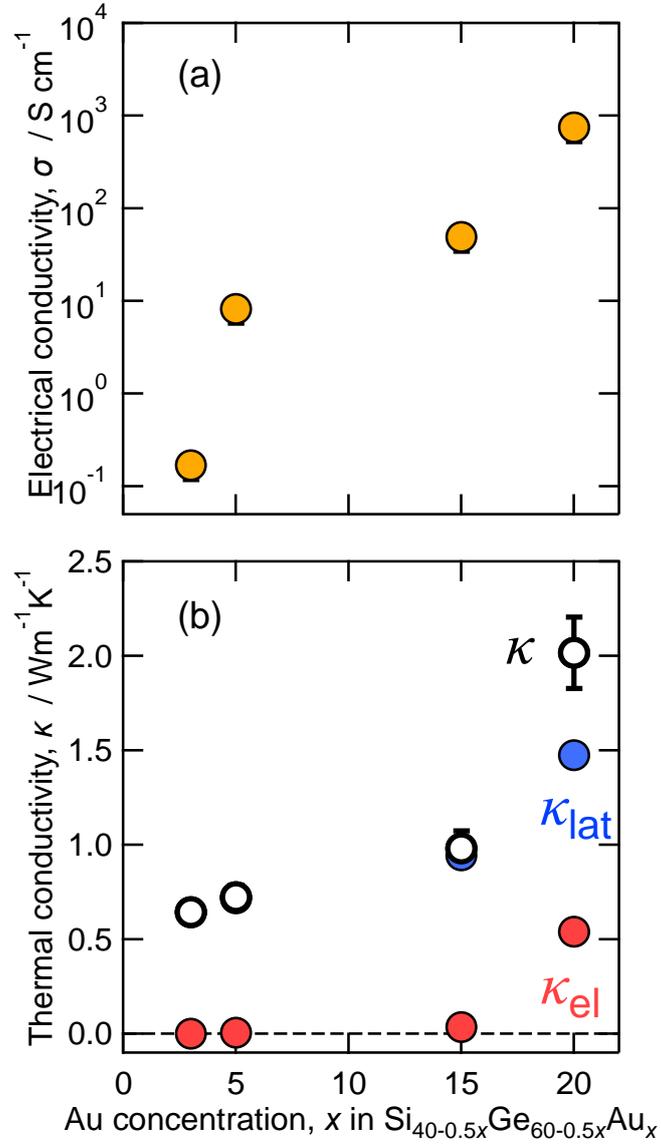

Fig. 5 (a) and (b) show the composition dependence of electrical conductivity and the electron ($\kappa_{el}$) and lattice vibration ($\kappa_{lat}$) contributions to the total thermal conductivity ($\kappa$) in for amorphous $Si_{40-x}Ge_{60-x}Au_x$ ($x$ = 3, 5, 25, and 20 at%) thin films, respectively.

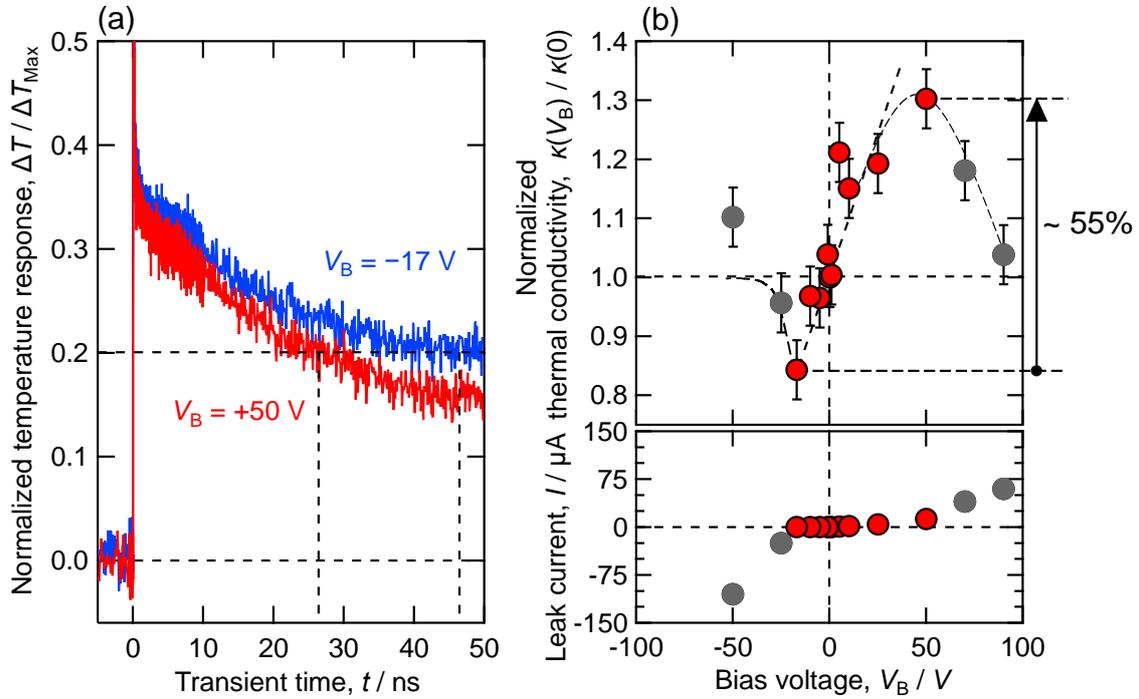

Fig. 6 (a) The transient curve of the normalized surface temperature of the Mo layer with a +50 V and –17 V bias voltage for which the largest heat flow density change was observed from the device shown in Fig. 2. (b) The bias voltage dependence of normalized "*thermal conductivity of the device*" and leak current through the device. The dashed line indicates a guide to the eye.